%% file: limits.tex
\documentclass[preprint,showpacs,preprintnumbers,amsmath,amssymb]{revtex4}

\def\bit{\begin{itemize}}
\def\eit{\end{itemize}}

\def\beq{\begin{equation}}
\def\eeq{\end{equation}}
\def\bea{\begin{eqnarray}}
\def\eea{\end{eqnarray}}
\usepackage{graphicx}
\usepackage{here}
\begin{document}

\preprint{FNAL-CONF-02/047-E}

\title{Confidence Limits and their Robustness}

\author{Rajendran Raja}
\email{raja@fnal.gov}
\affiliation{
Fermi National Accelerator laboratory\\
Batavia, IL 60510}

\date{\today}

\begin{abstract}
Confidence limits are common place in physics analysis. Great care
must be taken in their calculation and use, especially in cases of
limited statistics when often one-sided limits are quoted.
In order to estimate the stability of the
confidence levels to addition of more data and/or change of cuts, we argue
that the variance of their sampling distributions be calculated in
addition to the limit itself. The square root of the variance of their
sampling distribution can be thought of as a statistical error on the
limit.  We thus introduce the concept of statistical errors of confidence
limits and argue that not only should limits be calculated but also
their errors in order to represent the results of the analysis to the
fullest. We show that comparison of two different limits from two
different experiments becomes easier when their errors are also
quoted. Use of errors of confidence limits will lead to abatement of
the debate on which method is best suited to calculate confidence
limits.
\end{abstract}


\maketitle

\input introduction
\input motivate

\input algorithm

\end{document}

%% file: introduction.tex
\section{Introduction}

Confidence limits are used to express the results of experiments that
are not yet sensitive to discover the object of their searches. In
such cases, often a one-sided limit is used to delimit the quantity of
interest. Limits from different experiments are compared and attempts
are made to combine them. These limits can fluctuate up or down with
the addition of more data or the changing of the analysis parameters.
A measure of the robustness of the limits is given by the width of the
sampling distribution of these limits, where the sampling distribution
is obtained over an ensemble of similar experiments simulated by Monte
Carlo. The standard deviation of the sampling distribution of such
limits can be thought of as an error on the limit.
 
\begin{figure}[tbh!]
\centerline{\includegraphics[width=4.0in]{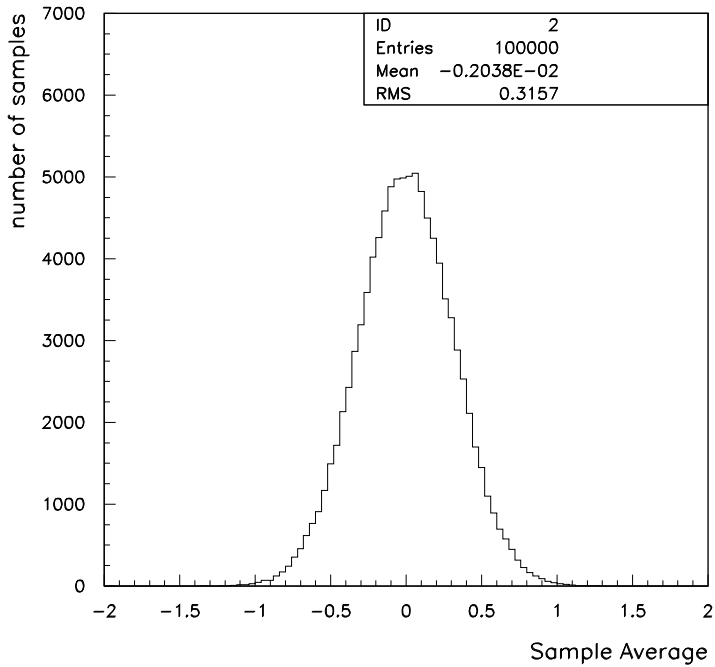}}
\caption[The average of the sample]
{The distribution of the sample average $\bar{x}$ over a large sample
of events.
\label{sample_average}}
\end{figure}

\begin{figure}[tbh!]
\centerline{\includegraphics[width=4.0in]{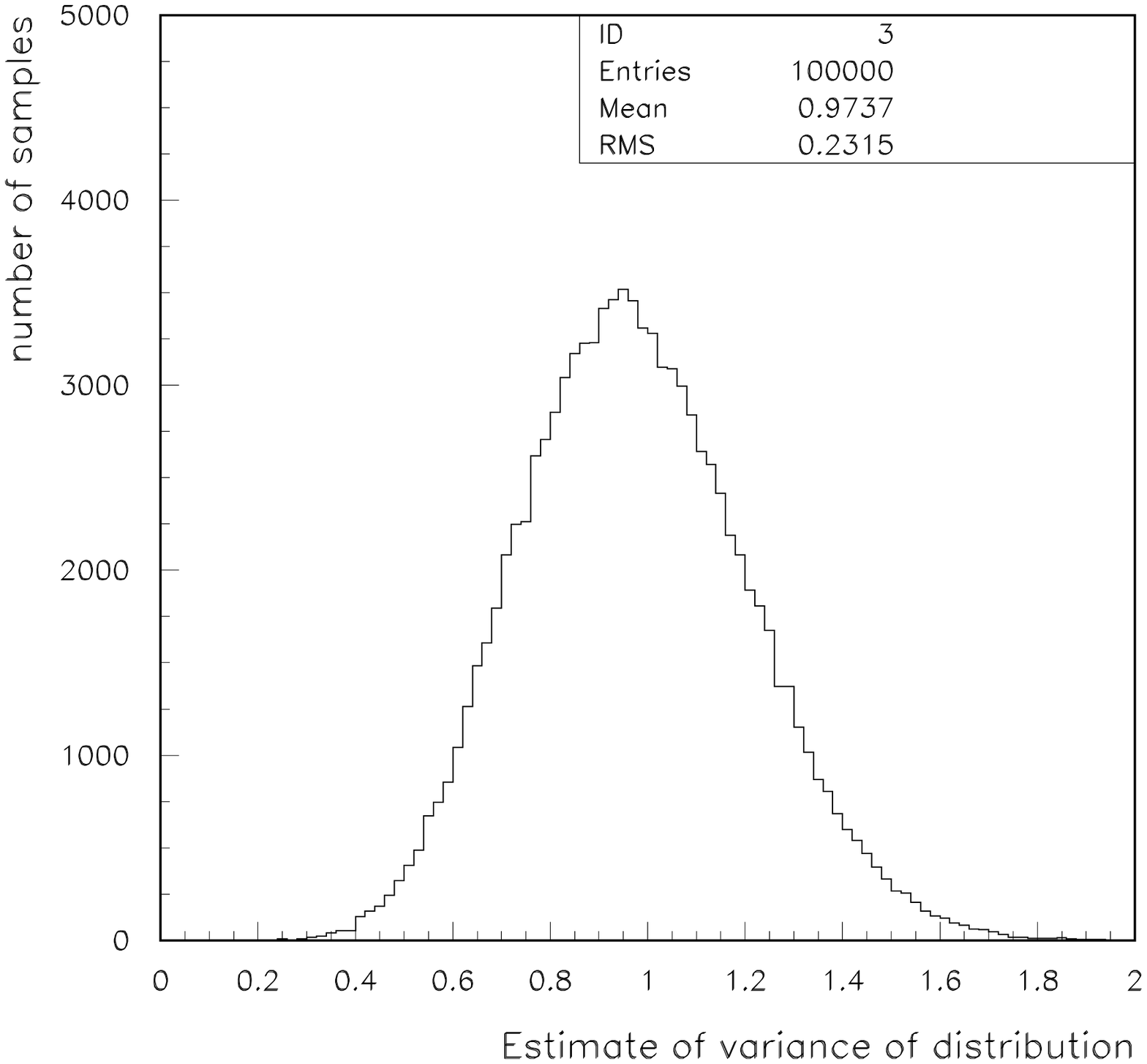}}
\caption[Unbiased estimate of the variance of the sample]
{Unbiased estimate $s$ of the standard deviation of the $\sigma$ of
the Gaussian distribution deduced from a sample of $n=10$ events. The
average value of $s$ is $\approx$ 1.0 and {\it its} standard deviation
is 0.23.
\label{var}}
\end{figure}

We introduce the concept of error of confidence limits by a simple
Gaussian example. Consider a sample of $n$ events, where $n=10$,
characterised by the variable $x$ distributed as a unit Gaussian, with
a mean value $\mu = 0$ and standard deviation $\sigma = 1$. Then the
average value $\bar{x}$ of the $n$ events will be distributed as a
Gaussian of mean value zero and standard error
$\sigma$/$\sqrt(n)$. The unbiased estimate of $\sigma$, the variance
of the distribution is given by $s$ where,
\begin{equation}
 s^2 = \frac{1}{n-1}\sum_{i=1}^{i=n}( x_i^2-\bar{x}^2)
\end{equation} 
Figure~\ref{sample_average} shows the distribution $\bar{x}$ of our
sample of 10 events for a large number of samples. The expected value
$\bar{x}$ is zero and its standard deviation is 0.32 which is
consistent with the theoretical value of $\sigma/\sqrt(n)$=0.316.
Figure~\ref{var} shows a histogram of $s$ deduced from a sample of 10
events for a large number of such samples. The average value of $s$ is
$\approx$ 1.0, showing that $s$ is an unbiased estimator of
$\sigma$. The important point to note is that $s$ {\it also} has a variance
and that  its standard deviation is $0.23$. This is as expected from theory
where the error on the standard deviation of a Gaussian sample~\cite{weather}
is
$\approx \sigma/\sqrt(2n)$=0.223.
Having got the value of $\bar{x}$ and $s$ for our sample, one can
proceed to work out confidence limits for our observation. The two-sided 
68$\%$ CL limits for our observation of $\bar{x}$ will be given
by the standard error $\sigma(\bar{x})$ of $\bar{x}$ and we would
write the observation of $\bar{x}$ from our sample as
\begin{equation}
   \bar{x} \pm \sigma(\bar{x}) = \bar{x} \pm s/\sqrt(n) = -0.188 \pm
   0.408
\label{err}
\end{equation}
where the numbers correspond to our sample of 10 events. Note that the
standard error $\sigma(\bar{x})$ = 0.408 derived from our sample of 10
events is quite different from the theoretical value of 0.32, but this
is merely due to statistical fluctuation.

One can also work out the two-sided 90\%~CL limits for our observation of
$\bar{x}$ which would correspond to $\pm 1.64~\sigma(\bar{x})$ and
quote the 90\%~CL limits as $-0.188 \pm 0.669$, 
which is the value observed for our sample of 10 events.

\begin{figure}[tbh!]
\centerline{\includegraphics[width=4.0in]{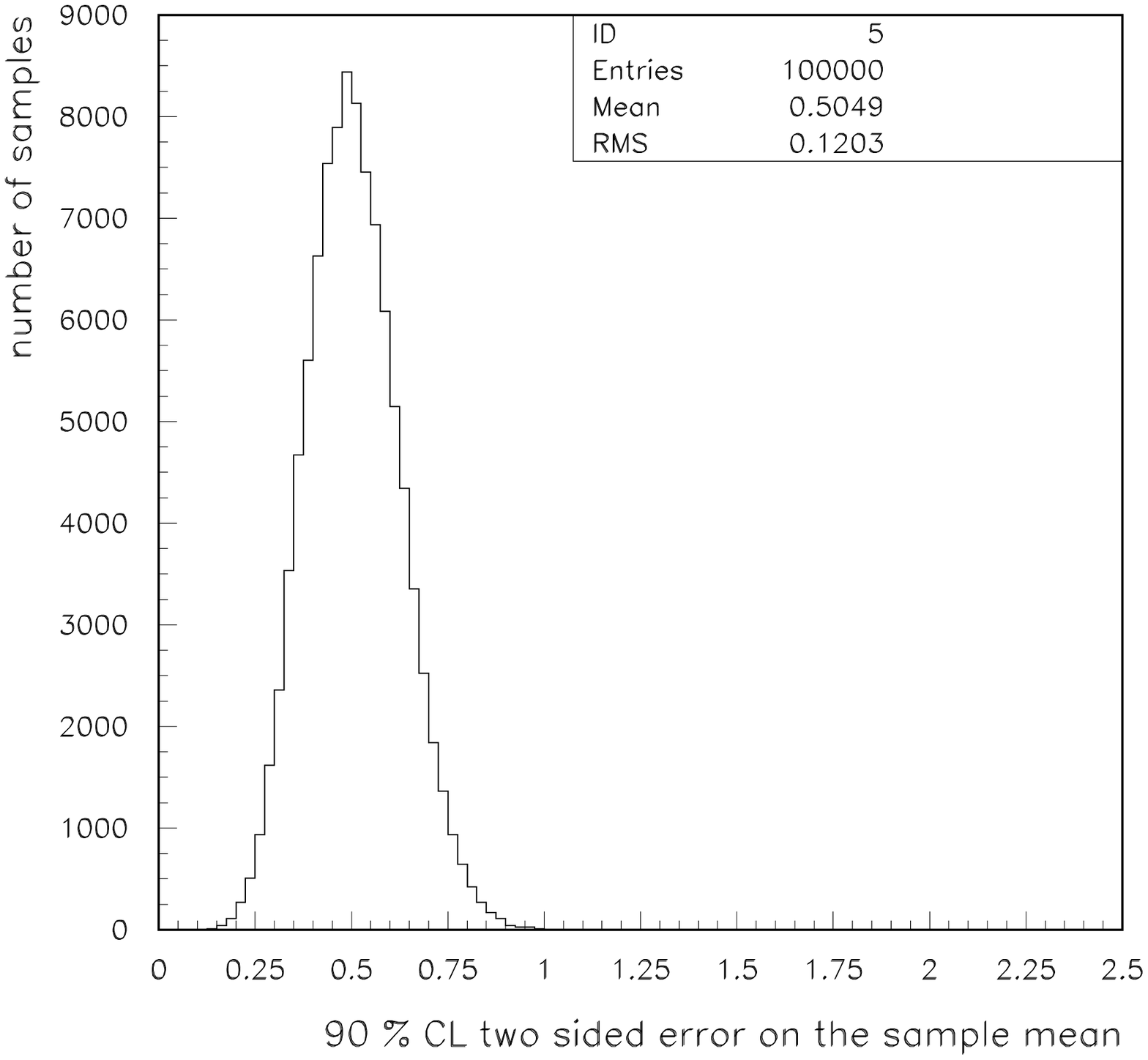}}
\caption[90\%~CL errors of the mean value ]
{The distribution of the calculated two-sided 90\%~CL errors of the
mean value of the sample.
\label{cl90}}
\end{figure}

\begin{figure}[tbh!]
\centerline{\includegraphics[width=4.0in]{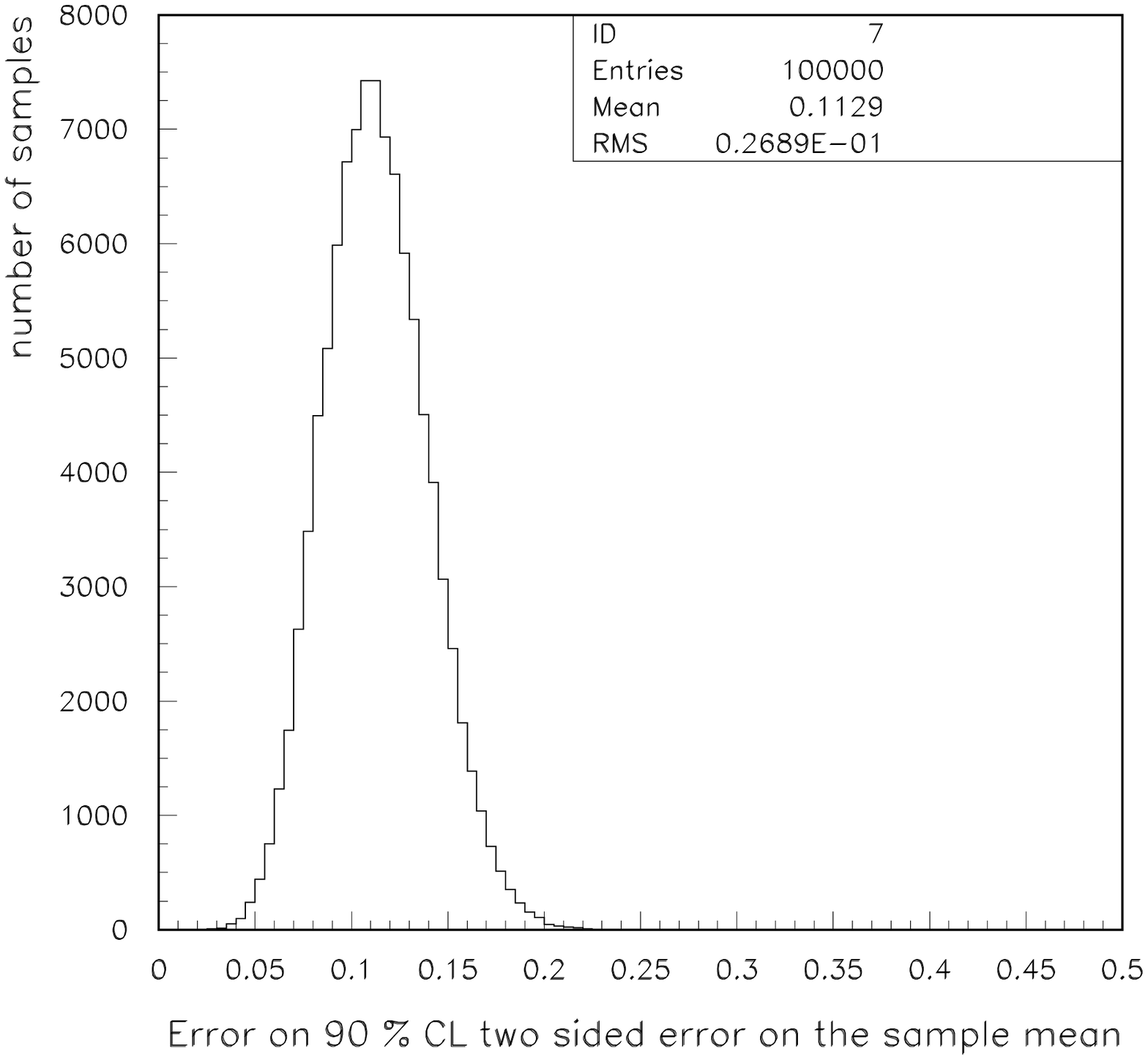}}
\caption[Error on the 90\%~CL errors of the mean value]
{The distribution of the calculated error on the two-sided 90\%~CL
error of the mean value of the sample.
\label{ecl90}}
\end{figure}

Figure~\ref{cl90} shows the distribution of the 90\%~CL two-sided
errors on the sample average,
over a large number of samples. The mean value of the
distribution is 0.505 which is close to the theoretical value of
1.64~$\sigma(\bar{x})$=0.519.  Note that the standard deviation of the
90\%~CL errors in Figure~\ref{cl90} is 0.12. We can also calculate the
standard deviation of the 90\%~CL error from our sample as 1.64
$\sigma(\bar{x})/\sqrt(2n)$ and this is plotted in figure~\ref{ecl90}.
The mean value of the standard deviation of the 90\%~CL error in 
figure~\ref{ecl90} is 0.113, in line with the theoretical value of 0.116.
 When the mean value is of
interest, we quote the mean value and the standard error on the mean
value as in equation~\ref{err}. This enables us to gauge the fluctuations in 
the mean value from sample to sample. When the confidence limit is of
interest, we propose that we quote the confidence limit along with
{\it its} standard error. This would enable us to gauge the significance 
and stability of the confidence level. In our example we would write this as
\begin{equation}
\bar{x} -1.64\sigma(\bar{x})\pm \sigma_{90} < \mu  < \bar{x} + 1.64\sigma(\bar{x})\pm \sigma_{90} ~ at~90\%~CL 
\label{lim1}
\end{equation}
where $\mu$ is the expectation value of $\bar{x}$ and the standard
error $\sigma_{90}$ on the 90\% CL limit would be given by
\begin{equation}
\sigma_{90} \approx \sigma(\bar{x})\sqrt{(1 + (1.64)^2/(2n)}
\label{err1}
\end{equation}
In our sample of 10 events, this would lead to
\begin{equation}
-0.857 \pm 0.434 < \mu < 0.481 \pm 0.434 ~at~90\%~CL
\end{equation}
Note that the error on the lower and upper 90\% CL limits are
correlated by the error on $\bar{x}$ which they have in common. Half
the difference between the lower and uper 90
\% CL limits is $1.64\sigma(\bar{x})$ and its error is 1.64$\sigma(\bar{x})/\sqrt(2n)$. 
These two errors added in quadrature yield the formula in
equation~\ref{err1}. The error in the 90\%~CL limit indicates to the
reader the stability of the limit and the statistical significance of
the result.

Very often, we are not interested in the mean value of our
observations but are more interested in the confidence limits, due to
the low statistics of the observation. We may only be interested in an
upper (one-sided) bound. So we would quote a 95\% CL upper bound on
$\mu$ as
\begin{equation}
 \mu < 0.481 \pm 0.434 ~at~95\%~CL
\end{equation}
A second sample of 10 events from the same distribution may yield a result
\begin{equation}
   \mu < 0.354 \pm 0.335~at~95\%~CL
\end{equation}
but we do not fall into the trap of declaring the second result a
better limit than the first, because both the limits are the same
within errors. If we did not quote the errors on the limits, we would
be tempted to declare the second limit superior to the first.

Similarly, as analyses proceed in discovery searches, events can go in
and out of samples, as cuts are refined and more data is
accumulated. Appearance of a single event in a sample can change the
confidence limit drastically, as was the case in the search for the
top quark. These changes can be understood as fluctuations of the
confidence limit within errors, if we were to quote not only the
confidence limit but also its error.
\section{Reconciliation with the Neyman definition of Confidence
limits} 
\begin{figure}[tbh!]
\centerline{\includegraphics[width=4.0in]{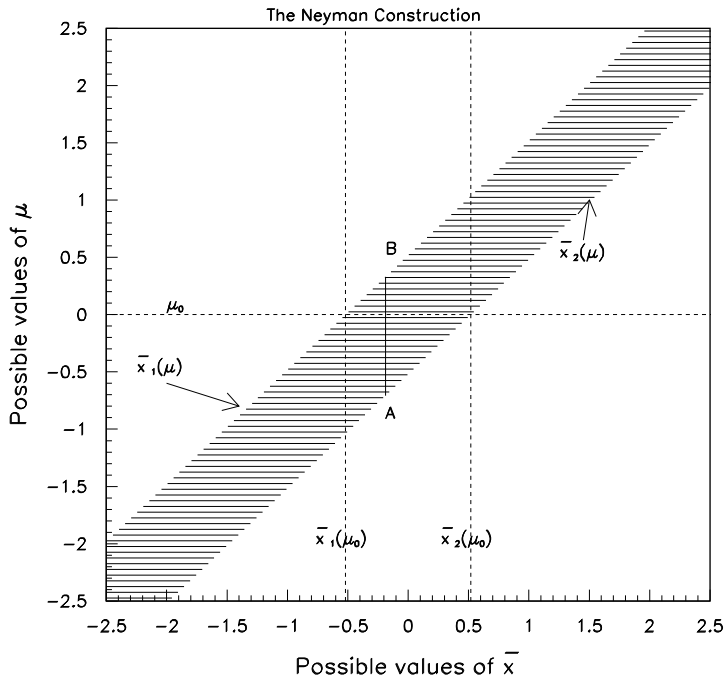}}
\caption[Neyman Construction of the confidence level]
{The Neyman construction of the confidence level for our example
\label{neym}}
\end{figure}

The construction of confidence levels as written down by
Neyman~\cite{neyman} may be understood within the context of our
current example as follows. Using our first sample of 10 events drawn
from a unit Gaussian, we calculate a mean value $\bar{x}=-0.188$. Let
us assume, for the sake of argument, that we know the variance of the
mean value to be $1.0/\sqrt(10)$. In this case, we can construct the
Neyman confidence level for $\mu$, the expectation value of $\bar{x}$,
as illustrated in Fig.~\ref{neym}. The parameter $\mu$ is plotted on the
ordinate and  $\bar{x}$  is plotted on the
abscissa. For each value of $\mu$, the 90\%~CL limits of $\bar{x}$ are
delineated by horizontal lines that are delimited by the curves
$\bar{x}_1(\mu)$ and $\bar{x}_2(\mu)$, assuming $\bar{x}$ is
distributed about $\mu$ with variance $1.0/\sqrt(10)$. If the true
value of $\mu$ is $\mu_0$, then $\bar{x}_1(\mu_0) < \bar{x}
<\bar{x}_2(\mu_0)$ with 90\% probability. If we now measure a value of
$\bar{x}=-0.188$, then we can construct the interval AB which will
contain the true value of $\mu_0$ if and only if $\bar{x}_1(\mu_0) <
\bar{x} <\bar{x}_2(\mu_0)$. In other words the interval AB has a
probability of 90\%~(also called ``coverage'') of containing the true
value $\mu_0$.  The interval AB is thus defined to be the 90\% CL
interval of $\mu$.

If we were however to repeat our measurement of $\bar{x}$ by creating
other samples of 10 events each, we would get different lines AB, each
of which would have a 90\% chance of containing the true value
$\mu_0$. Most of the time, one is interested in a central value of
$\bar{x}$ and an interval such as AB to denote the statistical errors
(robustness) of the measurement of $\bar{x}$. However, in experiments with poor
statistics, the central value $\bar{x}$ is often not of interest and the 
one-sided limit (either point A or B) is often quoted. At this stage, the
points A or B become point measurements in their own right, and it is
informative to quote their statistical errors in order to evaluate
their robustness.

\begin{figure}[tbh!]
\centerline{\includegraphics[width=4.0in]{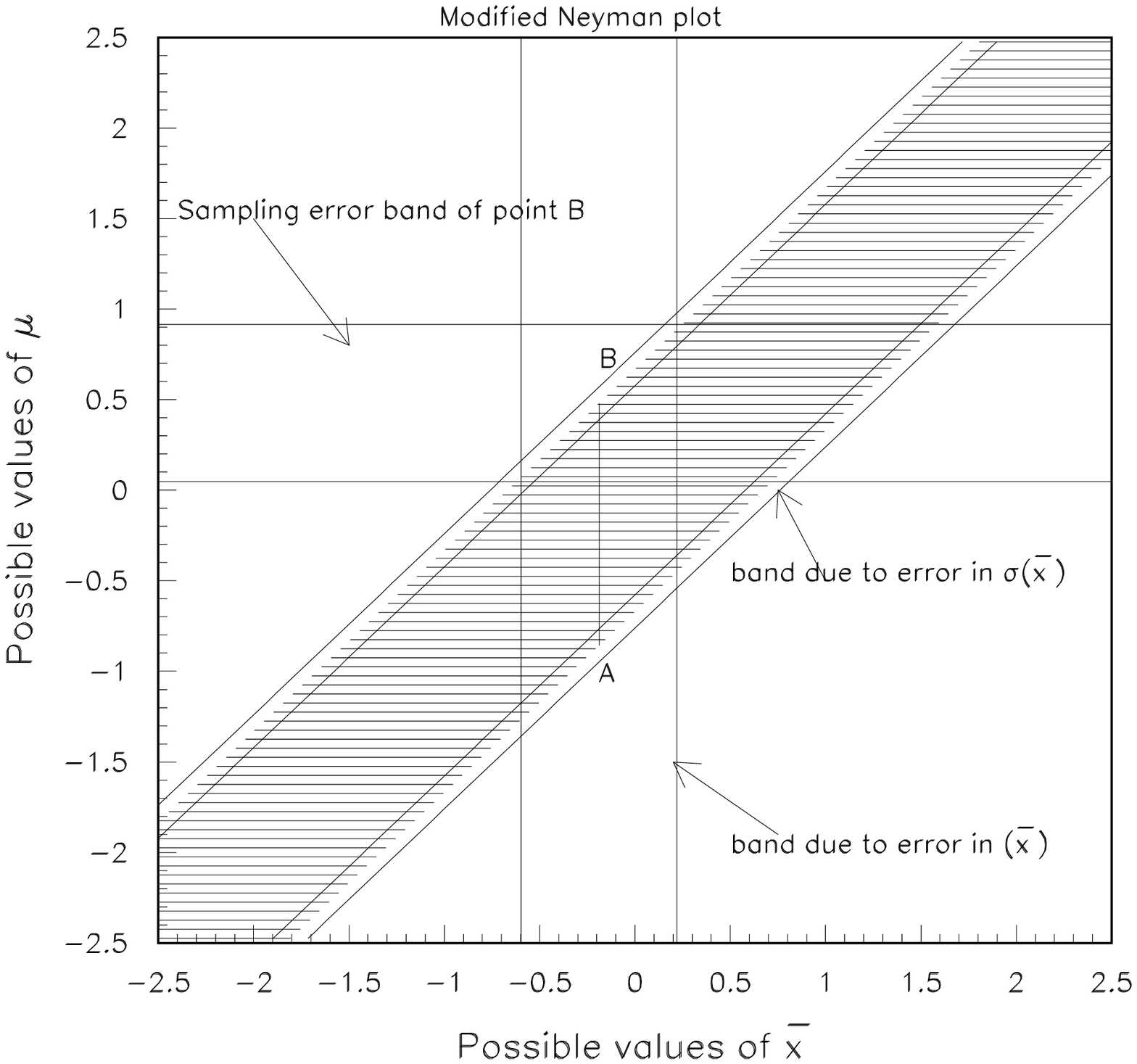}}
\caption[Modified Neyman Plot]
{The Neyman construction modified to illustrate fluctuations 
in $\bar{x}$ and $\sigma({\bar{x})}$ for our example. The error band 
due to $\sigma(\bar{x})$ and band due the error in $\sigma(\bar{x})$ are 
shown. These are added in quadrature to produce the sampling error 
band of point B.
\label{neym2}}
\end{figure}

This is illustrated further in Fig.~\ref{neym2}, where we now no
longer assume we know the variance of $\bar{x}$. This is computed from
the data and will fluctuate from sample to sample. These so-called
``nuisance variables'' are integrated over to yield a final confidence
limit in usual practice, which would be appropriate if one were
interested in the central value of $\bar{x}$. If however, one is
interested in the one-sided limit B, it would be appropriate to use
them to estimate the robustness of the point B due to statistical
fluctuations. We use the error bands shown for $\bar{x}$ and  $\sigma(\bar{x})$ in the figure to compute the sampling error band on the point B.


%% file: motivate.tex
\section{An Illustrative example} 
\begin{figure}[tbh!]
\centerline{\includegraphics[width=4.0in]{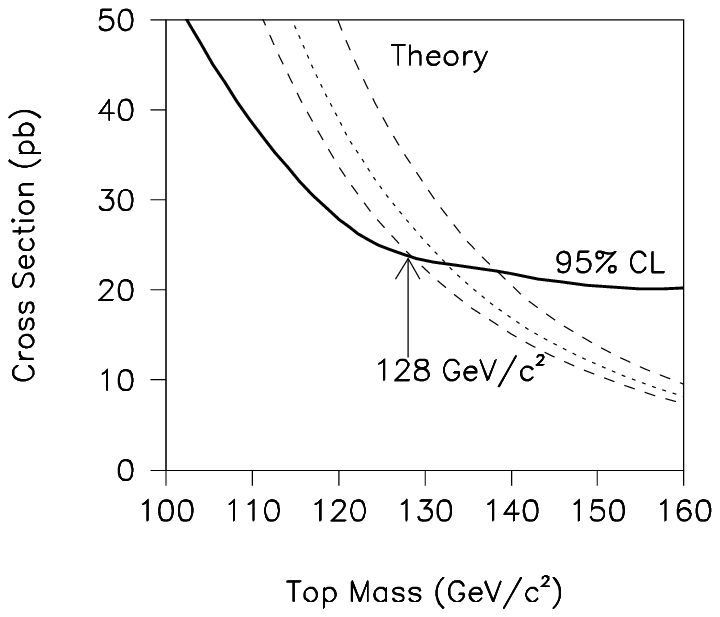}}
\caption[The D\O\ top quark 95\% CL upper limit]
{The 95\% confidence level~\cite{dzero} 
on $\sigma_{t\bar{t}}$ as a function of top
quark mass. Also shown are central (dotted line) and low (dashed line)
theoretical cross section curves~\cite{txsect}.
\label{d095}}
\end{figure}

\begin{figure}[tbh!]
\centerline{\includegraphics[width=4.0in]{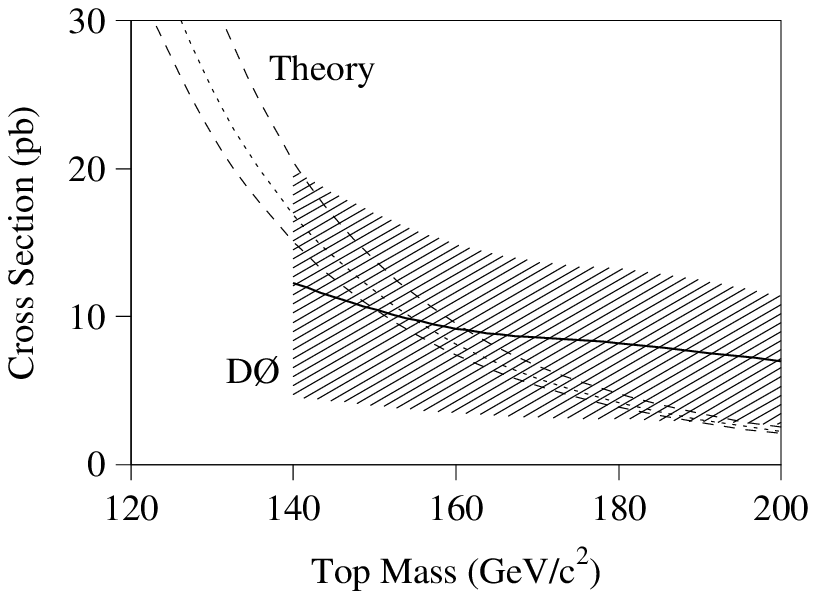}}
\caption[The D\O\ top quark cross section]
{Measured $t\bar{t}$ production cross section (solid line, shaded band
= one standard deviation error) as a function of top mass~\cite{dzero}. 
Also shown
are central (dotted line), high and low (dashed lines) theoretical
cross section curves~\cite{txsect}.
\label{d0xsect}}
\end{figure}

We can illustrate the need for confidence limits errors using the
following example. In 1995, the D\O\ collaboration published limits on
the top quark mass and cross section~\cite{dzero}. Figure~\ref{d095}
shows~\cite{dzero} the 95\% CL upper limit on top quark production as
a function of top quark mass using 13.5 pb$^{-1}$ of data. The
confidence limit curve is used to derive a lower limit of 128
GeV/c$^2$ for the top quark mass at 95\% CL. In the same paper,
another figure, reproduced here as Figure ~\ref{d0xsect} 
shows the top quark production
cross section as a function of the top quark mass. This curve has a
1~$\sigma$ error band around it. But the top quark production cross
section may be thought of as the 50\% CL upper/lower bound on the
cross section. Surely, if the 50\% CL limit has an error band around
it, the 95\% CL limit should also have its own error band.
In what follows, we show how to calculate errors in confidence levels
in general and use the method to calculate the error in the 95\% CL
curve shown in Figure~\ref{d095}.

%% file: algorithm.tex
\section{A general algorithm to calculate errors in Confidence Limits}

\begin{figure}[tbh!]
\centerline{\includegraphics[width=4.0in]{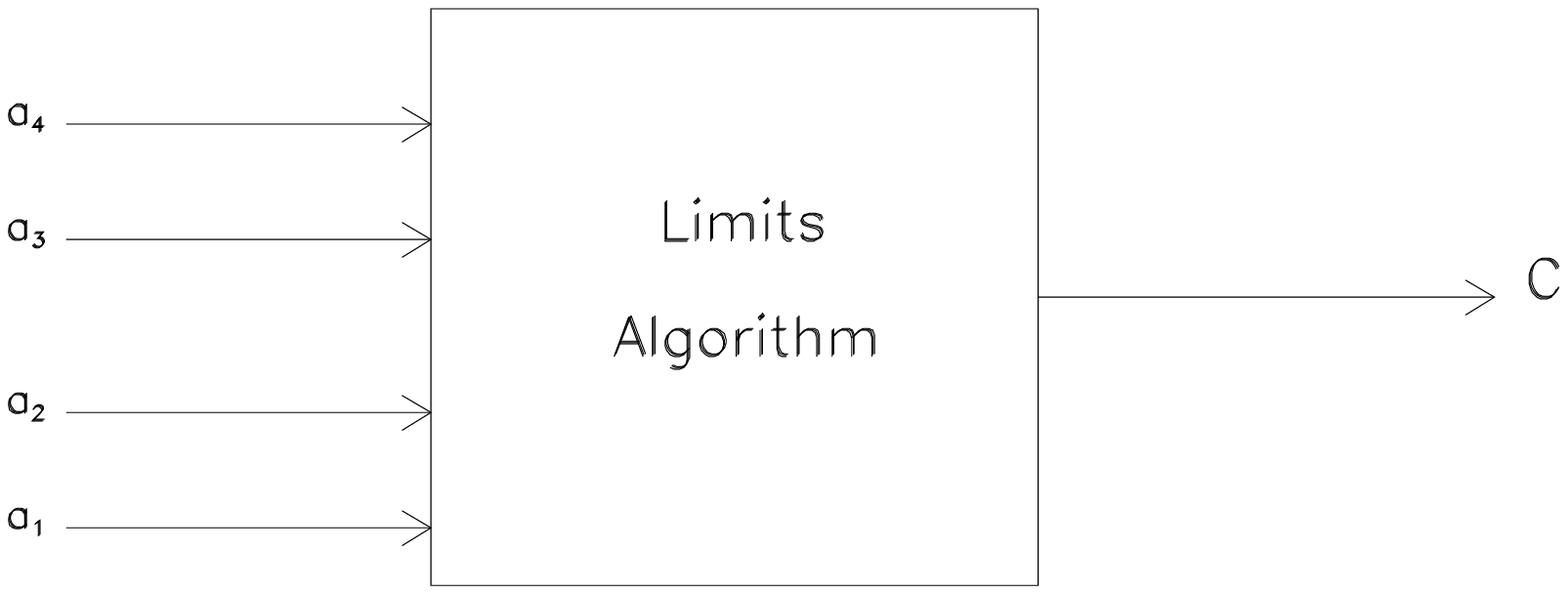}}
\caption[The limits algorithm]
{Schematic ``black box'' representation of a general confidence limit
calculating algorithm, that has input parameters $a_1,a_2..a_4$ and
outputs a confidence level $C$ in a single variable.
\label{box}}
\end{figure}

Most experiments have elaborate algorithms to calculate confidence
limits for their results. Such algorithms will include detailed
calculations and parametrizations of efficiencies and acceptances. In
addition, they will have several other input parameters such as the
number of events observed, total integrated luminosity and the
error on the luminosity. 
Let us denote the input parameters as $a_{i},~i=1,n$.
The output
of such a program will be the confidence limits
$C_{\alpha},~\alpha=1,k$. Figure~\ref{box} illustrates this general
case.
Then, for small changes in the input parameters, the following equations hold.
\begin{equation}
\delta C_{\alpha} = \frac{\delta C_\alpha}{\delta a_i} \delta a_i
\end{equation}
\begin{equation}
<\delta C_{\alpha}\delta C_{\beta}> = \frac{\delta C_\alpha}{\delta
a_i} \frac{\delta C_\beta}{\delta a_j}<\delta a_i \delta a_j>
\end{equation}
where the repeated indices $i,j$ are meant to be summed over and the
symbols $<>$ indicates the average over the enclosed quantities. The
quantity on the left hand side of the equation is the error matrix in
the confidence limits $C_\alpha$, denoted $E_{CC}$. The above equation
can be re-written in matrix form as
\begin{equation}
  E_{CC} = \tilde{T} E_{aa} T
\label{eqerr}
\end{equation}
where $E_{aa}$ is the error matrix of the input parameters $a_i,~i=1,n$
and $T$ is the transfer matrix, such that $T_{\alpha,i} = \frac{\delta
C_\alpha}{\delta a_i}$. $T$ can be determined numerically by varying
the input parameters to the limits algorithm. 
The error matrix $E_{aa}$ should be known to 
the experimenter, yielding the required error matrix $E_{CC}$.
\subsection{An Example}
Let us consider the calculation of $C$, the 95\% CL upper limit to the
top quark cross section as published in reference~\cite{dzero}. The
output of the limits algorithm is $C$. The input parameters can be
taken as three, namely $a_1$, the total number of top quark events observed,
$a_2$, the luminosity$\times$efficiency$\times$branching ratio of
the channels under consideration, summed over the channels and $a_3$,
the error in the luminosity. We have used a single parameter
$a_2$ summed over the channels to simplify the calculation. In
principle, all channels may be varied independently, but since they
are uncorrelated, and the dominant error is due to the common
luminosity factor, the above simplification will result.  We use this example
for illustrative purposes to show how such a calculation may proceed.

The error matrix of the parameters $E_{aa}$ is a 3$\times$3 diagonal
matrix, since the parameters are uncorrelated. The variance of $a_1$
is the number of events observed, the variance of $a_2$ is calculated
using the error in luminosity, and the variance of $a_3$ is calculated
assuming that there is a 50\% uncertainty in the {\it error} in the
luminosity. The transfer matrix $T$ is calculated by numerical
differentiation.

\begin{figure}[tbh!]
\centerline{\includegraphics[width=4.0in]{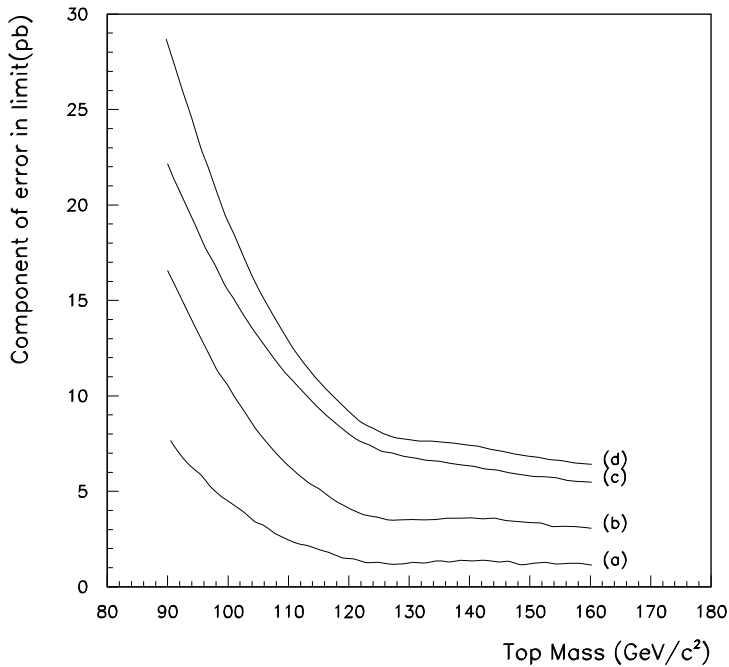}}
\caption[The component errors as a function of top mass]
{The components of $\sigma_C$,
the error in the 95\% CL top quark cross section upper limit,
 due to uncertainties in 
(a) error in luminosity 
(b) Luminosity$\times$efficiency$\times$branching ratio
(c) The overall number of events observed  as a function of top quark mass.
 (d) shows the overall error $\sigma_C$.

\label{toperr}}
\end{figure}

\begin{figure}[tbh]
\centerline{\includegraphics[width=4.0in]{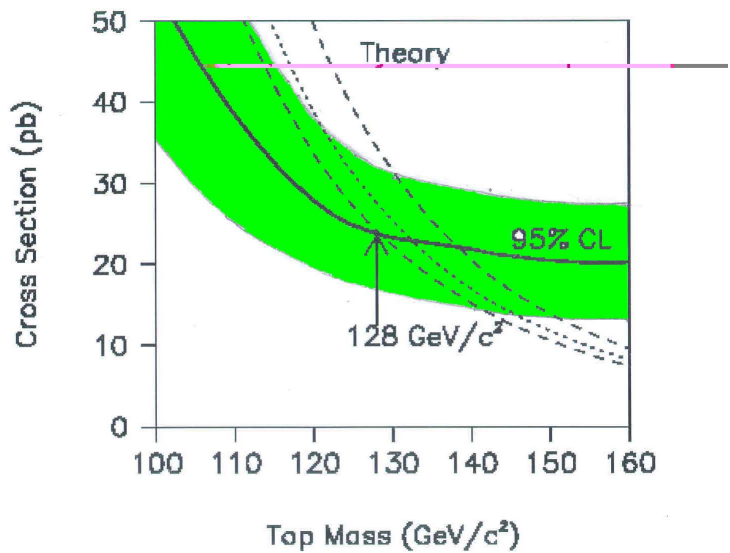}}
\caption[The D\O\ 95\% CL upper limit with errors]
{The D\O\ 95\% CL upper limit to the top quark cross section~\cite{dzero} with 
its accompanying error band, as calculated by the method in the text.
\label{d095lim}}
\end{figure}

Figure~\ref{toperr} shows the contribution to $\sigma_C$, the error in
the 95\%  CL upper limit to the cross section, due to the three
parameters $a_1$, $a_2$ and $a_3$ as a function of the top quark
mass. The overall error $\sigma_C$, obtained by adding the component
errors in quadrature, is also shown as a function of the top quark
mass. It can be seen that the contribution due to uncertainties in
$a_1$, is negligible. So we are not sensitive to errors in our guess 
of 50\% uncertainty to the error in the luminosity.  
The overall error  is dominated by the fluctuation in the
total number of events. This example thus graphically illustrates why
confidence limits fluctuate up and down as events fall in and out of
the selected sample as the analysis proceeds and more data is
accumulated. The 95\% CL upper limit to the cross section is merely
fluctuating within its error as all statistical quantities do.  When
we are interested in a confidence  limit, it thus behooves us to compute
not only that limit but also its error.
We may superimpose these errors on Figure~\ref{d095} yielding
Figure~\ref{d095lim}. The 95\% CL lower limit to the top quark mass
can then be quoted as 128$^{+14}_{-18}$~GeV/c$^2$, the error bars
indicating the range of fluctuation for the mass limit. This implies
that if one were to repeat the D\O\ experiment numerous times with an
integrated luminosity of 13.5 pb$^{-1}$ fluctuating within its errors,
one would expect to get a top quark lower mass limit that fluctuates
within the errors quoted.
\section{Combining limits}
Combining limits from two different experiments is difficult at best.
We remark here that in simple Gaussian cases, quoting the limit and
its error provides us with enough information to make a combined
result, as may be seen by examining equations \ref{lim1} and
\ref{err1}. Using the value of the limit and its error, we may deduce
$\bar{x}$ and $\sigma(\bar{x})$, if the number of events $n$ in the
sample is known.  Having the mean and its variance in each case, we
can combine the Gaussians, leading to a new variance for
the combined data. The combined mean of the two distributions can be
found as usual by the weighted average of the two means, the weights
being the inverse variances. It must be emphasized that the combined
limit is not simply the weighted average of the two limits as in the
case of the means. 

One can further ask if the two limits are consistent
with each other, if the errors on the limits are quoted, as shown below.
\section{Comparing Limits from two different algorithms}
When two different algorithms are used on the same data, two different
limits will result that are correlated. The correlations will be due
to the common input into the two algorithms. We can think of the
``black box'' in Fig.~\ref{box} as consisting of two different
algorithms producing as output $C_1$ and $C_2$, the two confidence
levels in question, using the same common input $a_{i},~i=1,n$.  We
can then use equation~\ref{eqerr} to work out $E_{CC}$, the error
matrix of the two confidence level algorithms and use this matrix to
decide whether the two confidence levels are significantly different
from each other as per,
\begin{equation}
 var (C_1-C_2) = var (C1) + var (C2) -2
 cov(C1,C2) = E_{11}+E{22}-2E_{12}
\end{equation}
\section{Conclusions} 
We have motivated the concept of statistical
error for a confidence limit, as the standard deviation of the
sampling distribution of such limits over an ensemble of similar
experiments.  In cases of limited statistics, our estimates of the
confidence limits can fluctuate significantly. Comparing confidence
limits becomes more meaningful when these errors are quoted. Different
methods exist (e.g Bayesian, Frequentist) for calculating these
limits. The differences between limits computed in the same experiment
using different methods will lose their significance if the limits are
shown to be the same within their sampling error.  Often in analyses
with limited statistics, the appearance of a new event can make
significant differences to the limit calculation. An error analysis of
the limit will show that the limit is exhibiting statistical
fluctuation as it is entitled to.  We propose that experimenters
publish confidence limits to their data accompanied by the error on
the limits.
\section{Acknowledgements}
The author wishes to thank Roger Barlow, Bob Cousins, Louis Lyons, 
Harrison Prosper, Byron Roe, and Tom Trippe for helpful comments.